\documentclass[letterpaper, nospread 
              ]{jacow}
\makeatletter%
	\ifboolexpr{bool{xetex}}
	 {\renewcommand{\Gin@extensions}{.pdf,%
	                    .png,.jpg,.bmp,.pict,.tif,.psd,.mac,.sga,.tga,.gif,%
	                    .eps,.ps,%
	                    }}{}
\makeatother

\ifboolexpr{bool{xetex} or bool{luatex}} 
 {}                                      
 {\usepackage[utf8]{inputenc}}           

\usepackage[USenglish]{babel}			 
\usepackage{graphicx}
\usepackage[final]{pdfpages}
\usepackage{multirow}
\usepackage{ragged2e}
\usepackage{dblfloatfix}
\usepackage{blindtext}

\usepackage{subfigure}
\usepackage{lipsum}

\ifboolexpr{bool{jacowbiblatex}}%
 {%
  \addbibresource{jacow-test.bib}
  \addbibresource{biblatex-examples.bib}
 }{}
\begin{document}

\title{A Non-parametric Density Estimation Approach to Measuring Beam Cooling in MICE\thanks{MICE has been made possible by grants from DOE, NSF (U.S.A), the INFN (Italy), the STFC (U.K.), the European Community under the European Commission Framework Programme 7, the Japan Society for the Promotion of Science and the Swiss National Science Foundation.}}

\author{Tanaz Angelina Mohayai\thanks{tmohayai@hawk.iit.edu}, Pavel Snopok, Illinois Institute of Technology, Chicago IL, USA \\  
David Neuffer, Fermilab, Batavia IL, USA \\
		on behalf of the MICE Collaboration
		}
	
\maketitle

\begin{abstract}
The goal of the international Muon Ionization Cooling Experiment (MICE) is to demonstrate muon beam ionization cooling for the first time. It constitutes a key part of the R\&D towards a future neutrino factory or muon collider. The intended MICE precision requires development of analysis tools that can account for any effects (e.g., optical aberrations) which may lead to inaccurate cooling measurements. Non-parametric density estimation techniques, in particular kernel density estimation (KDE), allow very precise calculations of the muon beam phase-space density and its increase as a result of cooling. In this study, kernel density estimation technique and its application to measuring the reduction in MICE muon beam phase-space volume is investigated.
\end{abstract}

\section{Muon Ionization Cooling Experiment}
The Muon Ionization Cooling Experiment (MICE) aims to demonstrate ionization cooling, the only beam cooling technique capable of reducing the muon beam phase-space volume within the short muon lifetime. In ionization cooling, the beam cooling occurs via ionization energy loss of muons in a low-$Z$ absorbing material.~A figure of merit for beam cooling is the measure of root-mean-square (RMS) emittance reduction. For optimum muon ionization cooling, the idea is to maximize the cooling effect from ionization energy loss and minimize the heating effect from multiple Coulomb scattering. These two effects are modeled in terms of the change in the normalized transverse RMS emittance~\cite{dave},
\begin{equation}\label{eq:cooling}
\frac{d\varepsilon_{\perp}}{dx} \approx -\frac{\varepsilon_{\perp} }{\beta ^{2}E_{\mu }}\left \langle \frac{dE}{dx} \right \rangle +\frac{\beta _{\perp} (13.6\textnormal{MeV}/c)^{2}} {2\beta ^{3}E_{\mu }m_{\mu }X_{0}},
\end{equation}
where $E_{\mu}$ is the muon energy, $\beta c$ is the muon velocity, $dE/dx$ is the magnitude of the ionization energy loss, $m_{\mu}$ is the muon mass, $X_{0}$ is the radiation length, and $\beta_{\perp}$ is the transverse beta function at the absorber. Setting the cooling term (first term of Eq.~\ref{eq:cooling}) equal to the heating term (second term of Eq.~\ref{eq:cooling}) leads to the the minimum achievable emittance, or equilibrium emittance\cite{dave}:
\begin{equation}\label{eq:equilibrium}
\varepsilon _{\perp } \cong \frac{\beta_{\perp }(13.6\textnormal{MeV}/c)^{2} }{2X_{0}\beta m_{\mu }}\left \langle \frac{dE}{dx} \right \rangle^{-1}.
\end{equation}
A smaller equilibrium emittance (compared to an input emittance) leads to a more effective emittance reduction, which from Eq.~\ref{eq:equilibrium} is achieved when the beta function, $\beta_{\perp}$ is minimized (strong focusing), and $X_{0}$ is maximized (low-$Z$ material) for a given energy loss, $dE/dx$~\cite{dave}. 
\begin{figure*}[!tbh]
\centering
\includegraphics*[width=0.8\textwidth]{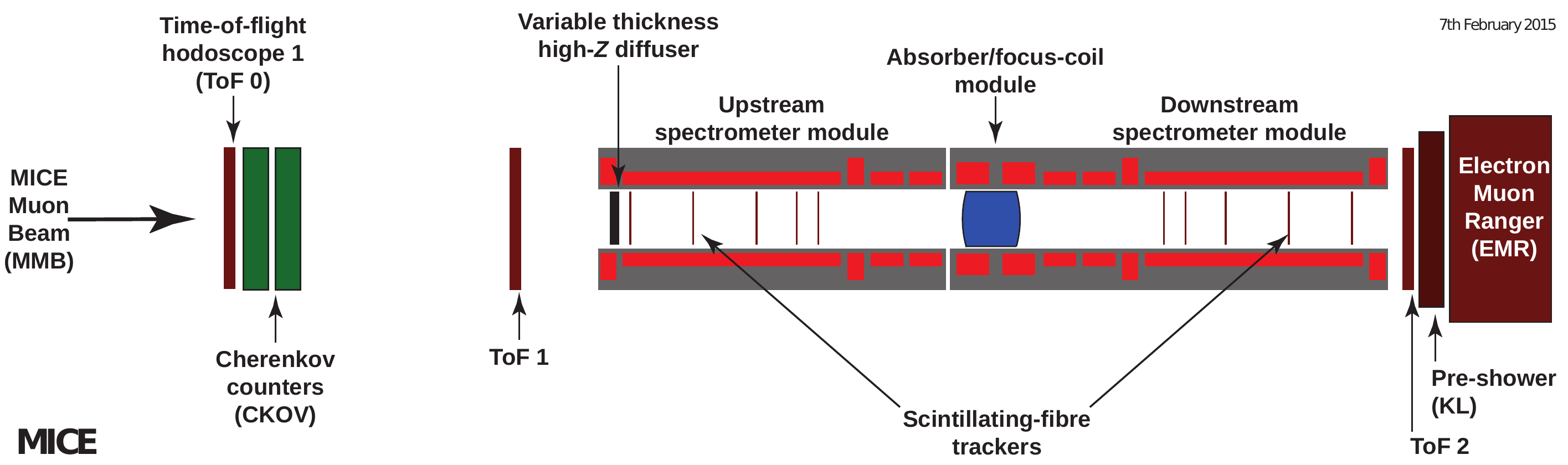}
    \caption{Schematic diagram of the Muon Ionization Cooling Experiment in its current experimental configuration.}
\label{fig:MICE}
\end{figure*}
In MICE, the cooling is measured for different optics configurations, input emittance values, and reference momentum values, and the measurement of each muon coordinate is done via the scintillating fiber tracking detectors. Each tracker is composed of five scintillating-fiber planar stations with three doublet fiber layers~\cite{tracker} immersed in the solenoidal fields of the Spectrometer Solenoids (SS) (Fig.~\ref{fig:MICE}). The upstream and downstream SS consist of five superconducting coils with two responsible for matching the MICE muon beam into and out of the absorber and three responsible for keeping the fields constant inside the tracking volumes. For measurement of beam cooling in MICE, the input and output beam distributions at the tracker stations immediately upstream and downstream of the absorber (tracker reference planes) are compared. Trackers reconstruct position and momentum coordinates of each muon through a series of cluster finding, space point reconstruction, and helical track fitting algorithms~\cite{tracker}. 

\section{Non-parametric Density Estimation}
\begin{figure}[h!]
    \centering
\setlength{\belowcaptionskip}{-10pt}
   \includegraphics[width=1\columnwidth]{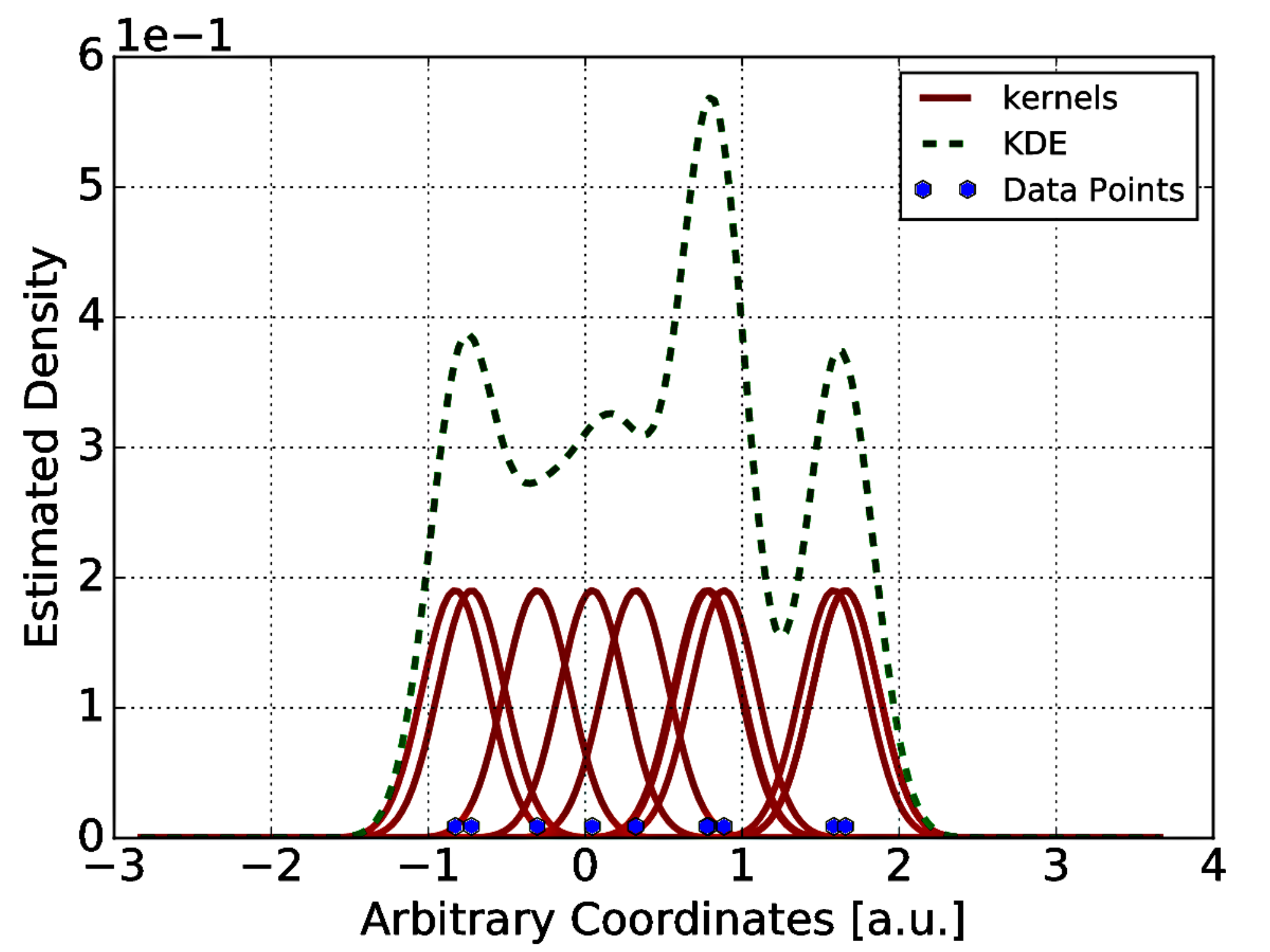}
   \caption{An illustration of the kernel density estimation technique.}
    \label{fig:kernels}
\end{figure}

The RMS emittance measurement can be sensitive to the tails of the beam distribution and can lead to inaccurate measures of beam cooling. Therefore, alternative measures of beam cooling such as the reduction in the beam phase-space volume and increase in phase-space density can be considered. MICE is a single particle measurement experiment that measures muon beam cooling to great precision, making it possible to use analysis tools that can estimate the underlying beam distribution more precisely.~Non-parametric density estimation techniques are examples of such analysis tools (first developed by Murray Rosenblatt in $1956$~\cite{Rosenblat}). The oldest and most well-known non-parametric density estimator is the histogram where the distribution function can be estimated via binning the data set and counting the data points inside each bin. 
\begin{figure}[h!]
    \centering
\vspace*{-\baselineskip}
\vspace{0.02mm}
   \includegraphics[width=1\columnwidth]{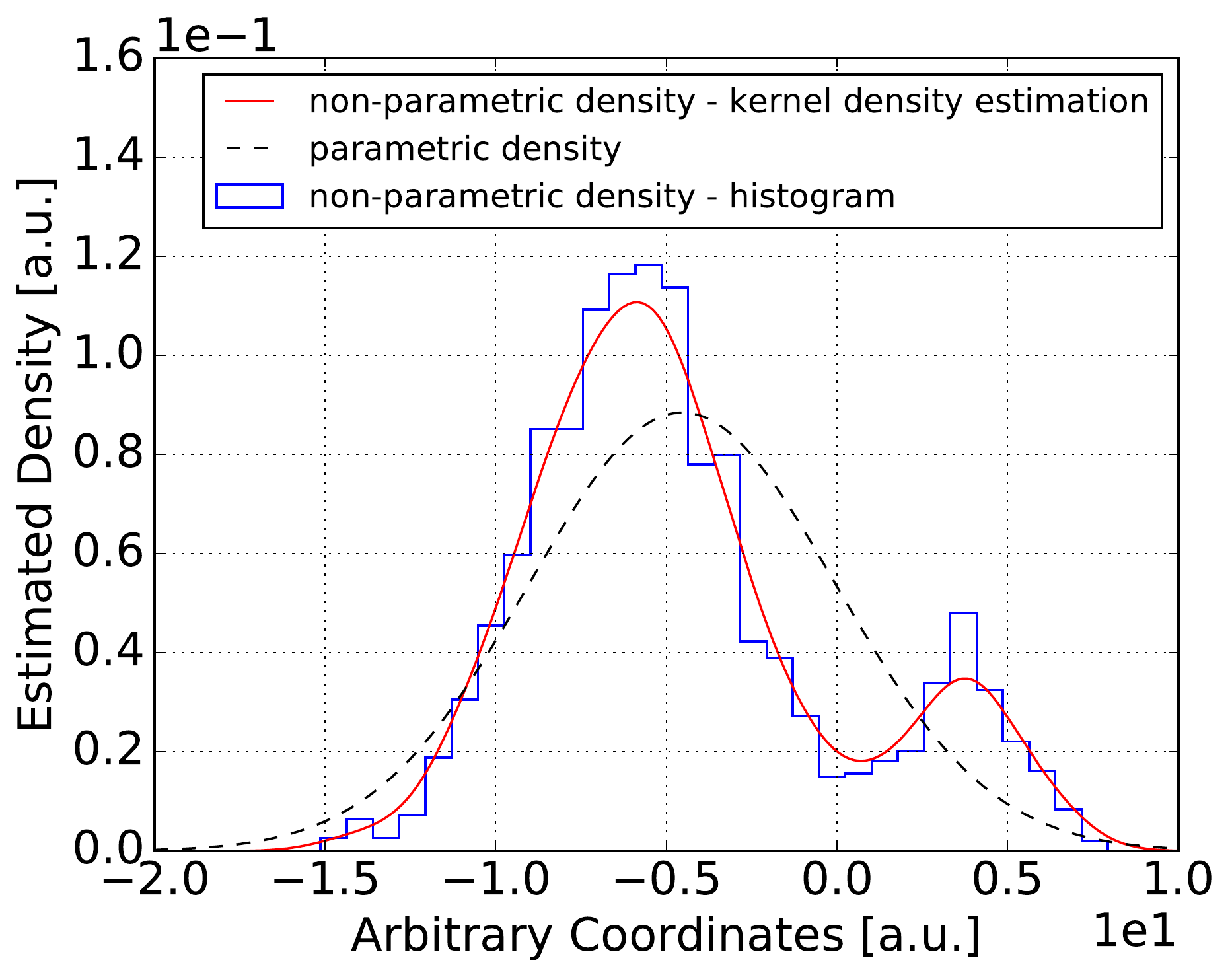}
   \caption{Comparison between the parametric density estimation approach and two non-parametric density estimation techniques.}
    \label{fig:comparison}
\setlength{\textfloatsep}{5pt}
\end{figure}
Determining the distribution function in this manner does not require a prior assumption about the functional form of the data. Parametric density estimation methods, on the other hand require an assumption about the underlying distribution; once a functional form for the distribution is assumed, the dataset is fit to that distribution for an estimation of the distribution parameters (e.g. distribution mean)~\cite{Rosenblat}. A more advanced non-parametric density estimator is the kernel density estimation (KDE) technique where instead of bins, smooth weight functions known as kernels are used~\cite{silverman} and $x, p_{x}, y, p_{y}$ of each muon is smeared with a Gaussian kernel. A comparison between the density curves estimated using the histogram, KDE, and the parametric approach is illustrated in Fig.~\ref{fig:comparison} where the underlying distribution is bimodal. The parametric approach assumes a Gaussian distribution and obscures the bimodal nature of the density curve. In addition, compared with the KDE approach, the histogram is a less smooth representation of the distribution. In the KDE approach, the density curve is obtained by summing over the kernels that are centered at each data point in the distribution~\cite{silverman}, 
\begin{equation}
\hat{f}(\vec{x})=\frac{1}{nh^{d}\sqrt{2\pi}}\sum_{i=1}^{n}k\left(\frac{-\left | \vec{x} - \vec{X}_{i} \right |^{2}}{2h^{2}}\right)
\end{equation}
where $k\left(\frac{-\left | \vec{x} - \vec{x_{i}} \right |^{2}}{2h^{2}}\right)$ is the kernel written as a function of reference point $\vec{x}$ (point at which the density function is evaluated), i$^{th}$ data point $\vec{X}_{i}$ (each i$^{th}$ data point contributes to the density at point $\vec{x}$ and represents the transverse position and momentum coordinates of each muon in MICE beam), and the width of the kernel $h$ (indicates the level of smoothing in the estimated density curve). $n$ is the sample size and $d$ is the dimensionality of the dataset. Fig.~\ref{fig:kernels} illustrates how the KDE approach estimates the underlying density function: the kernels are centered at each data point and are summed to estimate the underlying distribution~\cite{silverman}. 
KDE is therefore a powerful analysis tool for a single particle measurement experiment like MICE. In addition, the KDE compared to the histogram approach produces a higher resolution density measurement in higher dimensional phase space~\cite{silverman}, making it a more suitable alternative to histogram in transverse or six-dimensional beam cooling studies in MICE~\cite{silverman, KDE_1, KDE_2, KDE_3, KDE_4}. 
\section{Simulation Results}
\begin{figure}[tbh]
    \centering
\setlength{\belowcaptionskip}{-10pt}
   \includegraphics[width=1\columnwidth]{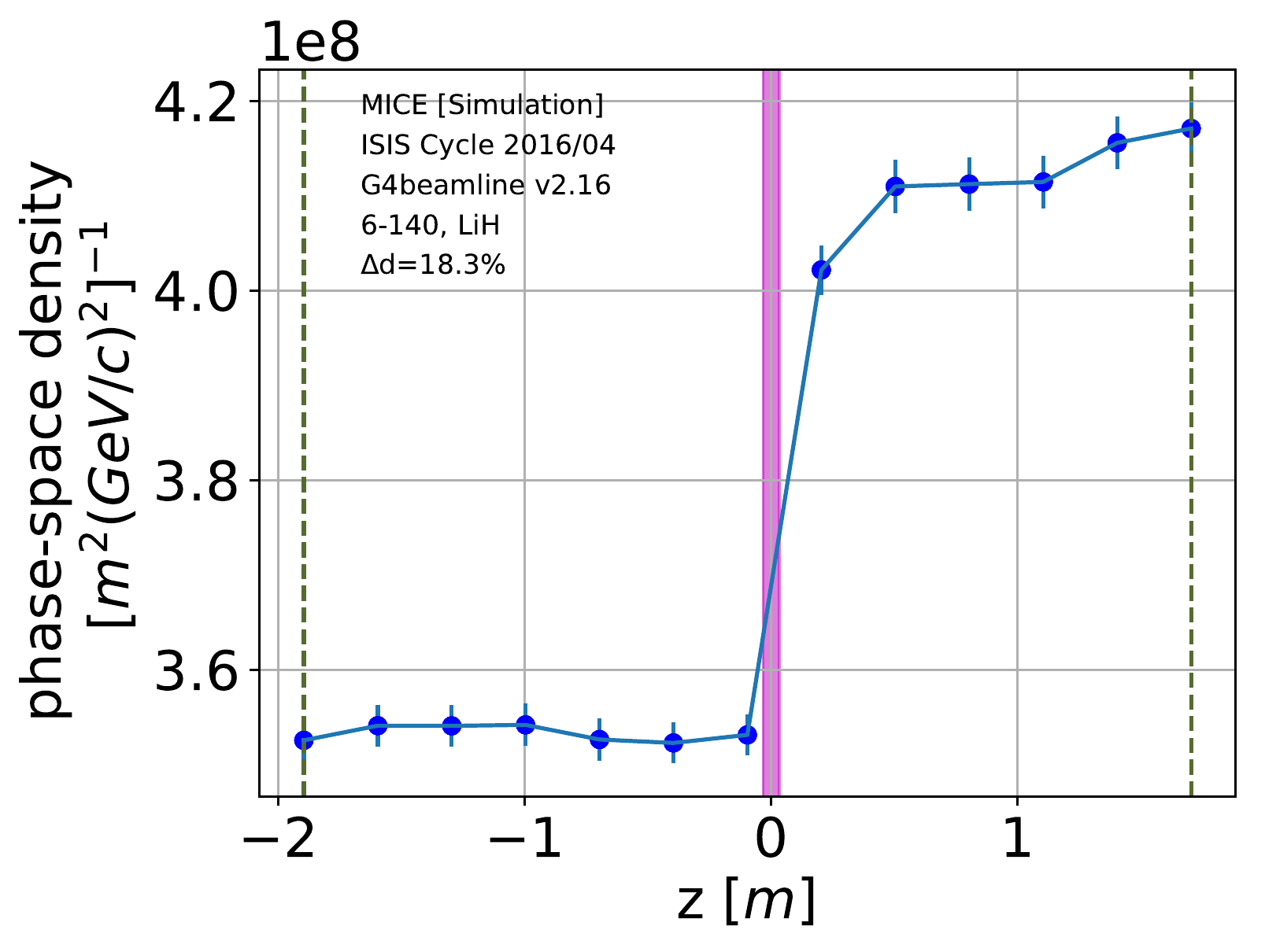}
   \caption{Evolution of the core phase-space density for the $6-140$ beam setting.}
    \label{fig:density}
\end{figure}
\begin{figure}[tbh]
    \centering
\setlength{\belowcaptionskip}{-10pt}
   \includegraphics[width=1\columnwidth]{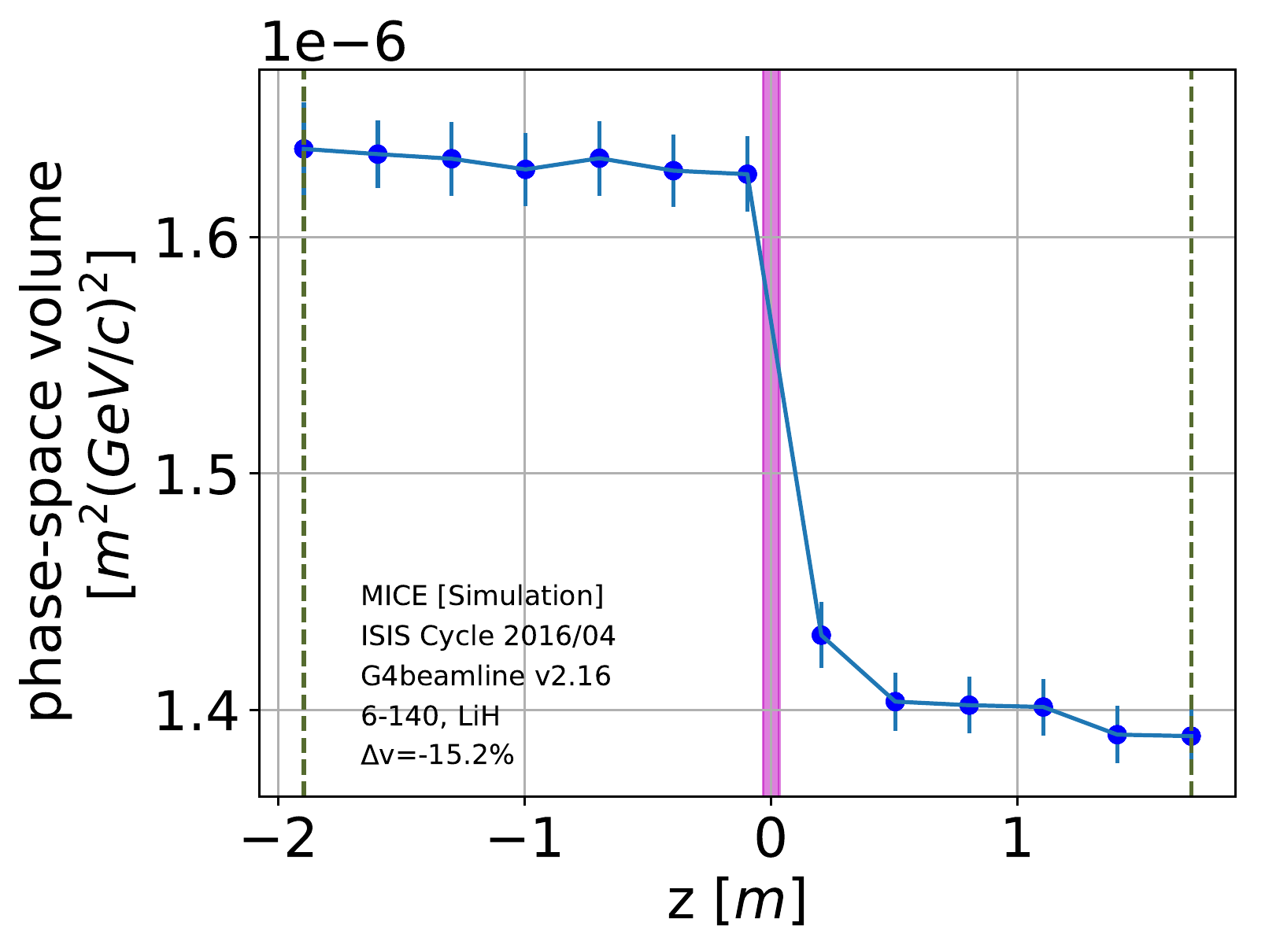}
   \caption{Evolution of the core phase-space volume for the $6-140$ beam setting.}
    \label{fig:volume}
\end{figure}
\begin{figure}[tbh]
    \centering
\setlength{\belowcaptionskip}{-10pt}
   \includegraphics[width=1\columnwidth]{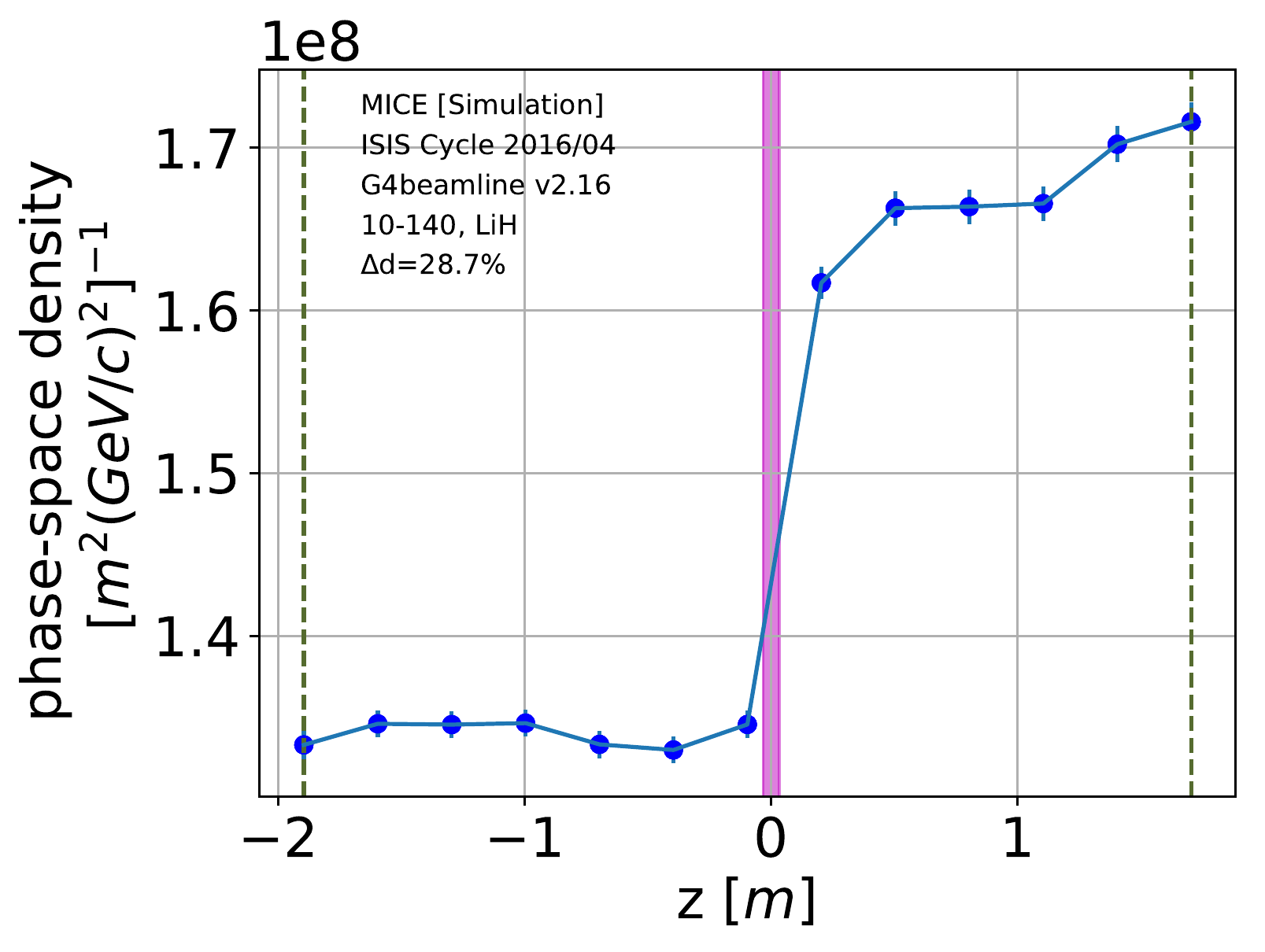}
   \caption{Evolution of the core phase-space density for the $10-140$ beam setting.}
    \label{fig:density_10}
\end{figure}
\begin{figure}[tbh]
    \centering
\setlength{\belowcaptionskip}{-10pt}
   \includegraphics[width=1\columnwidth]{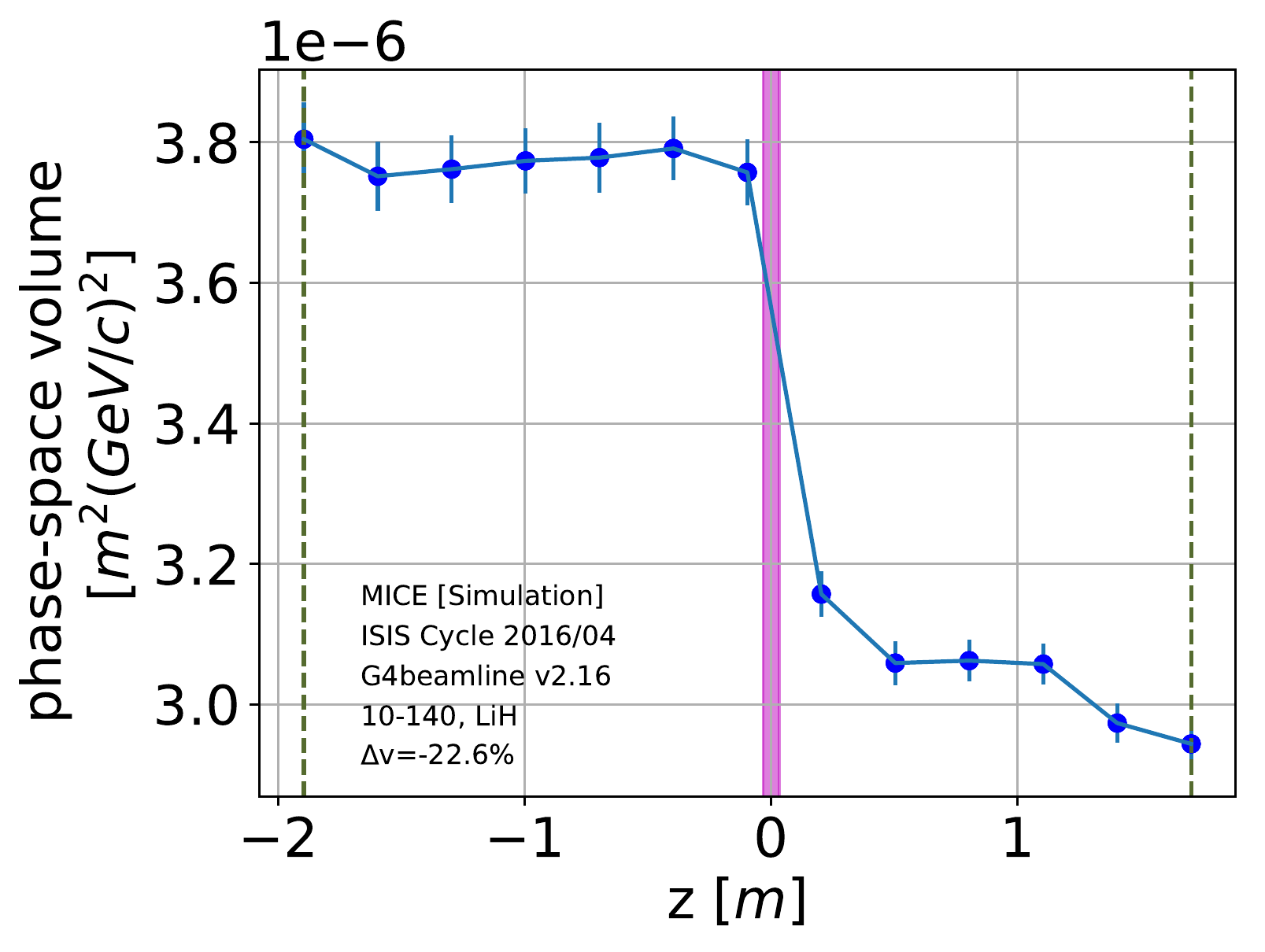}
   \caption{Evolution of the core phase-space volume for the $10-140$ beam setting.}
    \label{fig:volume_10}
\end{figure}
The MICE simulation studies in this paper are done using MAUS v3.0~\cite{MAUS} and G4beamline v2.16~\cite{g4beamline}. The MAUS routine generates the input muon distribution for tracking in G4beamline. The simulated magnet currents in the cooling channel are the same as the MICE runs from the user cycle 2016/04 (with two of the five coils in the downstream spectrometer solenoid turned off). The input beam contains a sample size of $10,000$ muons with a reference momentum of 140 MeV/$c$ and input normalized transverse RMS emittance values of $6$ mm (referred to as 6–140) and $10$ mm (referred to as 10-140). The cooling channel contains a LiH absorber which is centered at $z = 0$. There is particle loss due to scraping after passing through the LiH absorber. The simulated transmission efficiency is 85\%. No particle selection is applied downstream of the LiH absorber to discard scraped muons from the upstream sample.~Figures~\ref{fig:density},~\ref{fig:volume},~\ref{fig:density_10}, and~\ref{fig:volume_10} show the evolutions of the phase-space density and volume of the 9$^{th}$-percentile contour, representing the beam core (the contour enclosing a volume containing 9\% of the muon sample) along the MICE cooling channel for the $6-140$ and $10-140$ beam settings. The transverse phase-space coordinates are recorded along the z axis from location z $= -1.9$ m upstream of the absorber (upstream tracker reference plane) to z $= 1.9$ m downstream of the absorber (downstream tracker reference plane). The phase-space density and volume respectively increase and decrease as a result of beam cooling. The volume of the $9^{\textnormal{th}}$-percentile contour is obtained using the Monte Carlo (MC) volume calculation method~\cite{KDE_1, KDE_2, KDE_3, KDE_4}. Figures~\ref{fig:density_10} and~\ref{fig:volume_10} (a simulation study with an input emittance of $10$ mm) demonstrate larger percentage changes in density and volume compared with Fig.~\ref{fig:density} and Fig.~\ref{fig:volume}; this is because the input emittance of the beam is much larger ($10$ mm) for the same magnet setting and equilibrium emittance as the $6-140$ case.
\section{Conclusion}
The KDE technique has been applied to MICE simulation to characterize the MICE muon beam. KDE is a powerful non-parametric analysis approach in predicting the underlying density function of an input data set without any assumptions of the distribution functional form. KDE has also been used to isolate the core of the distribution where the beam distribution is free from any tail effects.
\section{Acknowledgement}
The first author acknowledges the support that she has received from United States National Science Foundation, the Division of Physics of Beams of the American Physical Society, and TRIUMF to attend IPAC 2018.

\end{document}